\documentclass[longauth]{aa}

\usepackage{txfonts}
\usepackage{graphicx}
\usepackage{mathtools}
\usepackage{float}
\usepackage{epstopdf}
\usepackage{amsmath}

\bibpunct{(}{)}{;}{a}{}{,} 

\begin{document}

\title{LOFAR tied-array imaging of Type III solar radio bursts}

\author{D.~E.~Morosan\inst{\ref{1}} 
 \and P.~T.~Gallagher\inst{\ref{1}} 
 \and P.~Zucca\inst{\ref{1}} 
 \and R.~Fallows\inst{\ref{2}} 
 \and E.~P.~Carley\inst{\ref{1}} 
 \and G.~Mann\inst{\ref{3}} 
 \and M.~M.~Bisi\inst{\ref{4}} 
 \and A.~Kerdraon\inst{\ref{5}} 
 \and A.~A.~Konovalenko\inst{\ref{6}} 
 \and A.~L.~MacKinnon\inst{\ref{7}} 
 \and H.~O.~Rucker\inst{\ref{8}} 
 \and B.~Thid\'e\inst{\ref{9}} 
 \and J.~Magdaleni\'c\inst{\ref{10}} 
 \and C.~Vocks\inst{\ref{3}} 
 \and H.~Reid\inst{\ref{7}} 
 \and J.~Anderson\inst{\ref{3}} 
 \and A.~Asgekar\inst{\ref{2}} 
 \and\inst{\ref{11}} 
 \and I.~M.~Avruch\inst{\ref{12}} 
 \and\inst{\ref{13}} 
 \and M.~J.~Bentum\inst{\ref{2}} 
 \and G.~Bernardi\inst{\ref{14}} 
 \and P.~Best\inst{\ref{15}} 
 \and A.~Bonafede\inst{\ref{16}} 
 \and J.~Bregman\inst{\ref{2}} 
 \and F.~Breitling\inst{\ref{3}} 
 \and J.~Broderick\inst{\ref{17}} 
 \and M.~Br\"uggen\inst{\ref{16}} 
 \and H.~R.~Butcher\inst{\ref{18}} 
 \and B.~Ciardi\inst{\ref{19}} 
 \and J.~E.~Conway\inst{\ref{20}} 
 \and F.~de Gasperin\inst{\ref{16}} 
 \and E.~de Geus\inst{\ref{2}} 
 \and A.~Deller\inst{\ref{2}} 
 \and S.~Duscha\inst{\ref{2}} 
 \and J.~Eisl\"offel\inst{\ref{21}} 
 \and D.~Engels\inst{\ref{22}} 
 \and H.~Falcke\inst{\ref{23}} 
 \and\inst{\ref{2}} 
 \and C.~Ferrari\inst{\ref{24}} 
 \and W.~Frieswijk\inst{\ref{2}} 
 \and M.~A.~Garrett\inst{\ref{2}} 
 \and\inst{\ref{25}} 
 \and J.~Grie\ss{}meier\inst{\ref{26}} 
 \and\inst{\ref{27}} 
 \and A.~W.~Gunst\inst{\ref{2}} 
 \and T.~E.~Hassall\inst{\ref{17}} 
 \and\inst{\ref{28}} 
 \and J.~W.~T.~Hessels\inst{\ref{2}} 
 \and\inst{\ref{29}} 
 \and M.~Hoeft\inst{\ref{21}} 
 \and J.~H\"orandel\inst{\ref{23}} 
 \and A.~Horneffer\inst{\ref{30}} 
 \and M.~Iacobelli\inst{\ref{25}} 
 \and E.~Juette\inst{\ref{31}} 
 \and A. ~Karastergiou\inst{\ref{32}} 
 \and V.~I.~Kondratiev\inst{\ref{2}} 
 \and\inst{\ref{33}} 
 \and M.~Kramer\inst{\ref{30}} 
 \and\inst{\ref{28}} 
 \and M.~Kuniyoshi\inst{\ref{30}} 
 \and G.~Kuper\inst{\ref{2}} 
 \and P.~Maat\inst{\ref{2}} 
 \and S.~Markoff\inst{\ref{29}} 
 \and J.~P.~McKean\inst{\ref{2}} 
 \and D.~D.~Mulcahy\inst{\ref{30}} 
 \and H.~Munk\inst{\ref{2}} 
 \and A.~Nelles\inst{\ref{23}} 
 \and M.~J.~Norden\inst{\ref{2}} 
 \and E.~Orru\inst{\ref{2}} 
 \and H.~Paas\inst{\ref{34}} 
 \and M.~Pandey-Pommier\inst{\ref{35}} 
 \and V.~N.~Pandey\inst{\ref{2}} 
 \and G.~Pietka\inst{\ref{32}} 
 \and R.~Pizzo\inst{\ref{2}} 
 \and A.~G.~Polatidis\inst{\ref{2}} 
 \and W.~Reich\inst{\ref{30}} 
 \and H.~R\"ottgering\inst{\ref{25}} 
 \and A.~M.~M.~Scaife\inst{\ref{17}} 
 \and D.~Schwarz\inst{\ref{36}} 
 \and M.~Serylak\inst{\ref{32}} 
 \and O.~Smirnov\inst{\ref{37}} 
 \and\inst{\ref{38}} 
 \and B.~W.~Stappers\inst{\ref{28}} 
 \and A.~Stewart\inst{\ref{32}} 
 \and M.~Tagger\inst{\ref{26}} 
 \and Y.~Tang\inst{\ref{2}} 
 \and C.~Tasse\inst{\ref{5}} 
 \and S.~Thoudam\inst{\ref{23}} 
 \and C.~Toribio\inst{\ref{2}} 
 \and R.~Vermeulen\inst{\ref{2}} 
 \and R.~J.~van Weeren\inst{\ref{14}} 
 \and O.~Wucknitz\inst{\ref{39}} 
 \and\inst{\ref{30}} 
 \and S.~Yatawatta\inst{\ref{2}} 
 \and P.~Zarka\inst{\ref{5}}
 }

\institute{ School of Physics, Trinity College Dublin, Dublin 2, Ireland \label{1}
 \and
Netherlands Institute for Radio Astronomy (ASTRON), Postbus 2, 7990 AA Dwingeloo, The Netherlands \label{2}
 \and
Leibniz-Institut f\"{u}r Astrophysik Potsdam (AIP), An der Sternwarte 16, 14482 Potsdam, Germany \label{3}
 \and
RAL Space, Science and Technology Facilities Council, Rutherford Appleton Laboratory, Harwell Oxford, Oxfordshire, OX11 OQX, United Kingdom \label{4}
 \and
LESIA, UMR CNRS 8109, Observatoire de Paris, 92195   Meudon, France \label{5}
 \and
Institute of Radio Astronomy, 4, Chervonopraporna Str., 61002 Kharkiv, Ukraine \label{6}
 \and
School of Physics and Astronomy, SUPA, University of Glasgow, Glasgow G12 8QQ, United Kingdom \label{7}
 \and
Space Research Institute, Austrian Academy of Sciences, Schmiedlstrasse 6, 8042 Graz, Austria \label{8}
 \and
Swedish Institute of Space Physics, Box 537, SE-75121 Uppsala, Sweden \label{9}
 \and
Solar-Terrestrial Center of Excellence, SIDC, Royal Observatory of Belgium, Avenue Circulaire 3, B-1180 Brussels, Belgium \label{10}
 \and
Shell Technology Center, Bangalore, India \label{11}
 \and
SRON Netherlands Insitute for Space Research, PO Box 800, 9700 AV Groningen, The Netherlands \label{12}
 \and
Kapteyn Astronomical Institute, PO Box 800, 9700 AV Groningen, The Netherlands \label{13}
 \and
Harvard-Smithsonian Center for Astrophysics, 60 Garden Street, Cambridge, MA 02138, USA \label{14}
 \and
Institute for Astronomy, University of Edinburgh, Royal Observatory of Edinburgh, Blackford Hill, Edinburgh EH9 3HJ, UK \label{15}
 \and
University of Hamburg, Gojenbergsweg 112, 21029 Hamburg, Germany \label{16}
 \and
School of Physics and Astronomy, University of Southampton, Southampton, SO17 1BJ, UK \label{17}
 \and
Research School of Astronomy and Astrophysics, Australian National University, Mt Stromlo Obs., via Cotter Road, Weston, A.C.T. 2611, Australia \label{18}
 \and
Max Planck Institute for Astrophysics, Karl Schwarzschild Str. 1, 85741 Garching, Germany \label{19}
 \and
Onsala Space Observatory, Dept. of Earth and Space Sciences, Chalmers University of Technology, SE-43992 Onsala, Sweden \label{20}
 \and
Th\"{u}ringer Landessternwarte, Sternwarte 5, D-07778 Tautenburg, Germany \label{21}
 \and
Hamburger Sternwarte, Gojenbergsweg 112, D-21029 Hamburg \label{22}
 \and
Department of Astrophysics/IMAPP, Radboud University Nijmegen, P.O. Box 9010, 6500 GL Nijmegen, The Netherlands \label{23}
 \and
Laboratoire Lagrange, UMR7293, Universit\`{e} de Nice Sophia-Antipolis, CNRS, Observatoire de la C\'{o}te d'Azur, 06300 Nice, France \label{24}
 \and
Leiden Observatory, Leiden University, PO Box 9513, 2300 RA Leiden, The Netherlands \label{25}
 \and
LPC2E - Universite d'Orleans/CNRS \label{26}
 \and
Station de Radioastronomie de Nancay, Observatoire de Paris - CNRS/INSU, USR 704 - Univ. Orleans, OSUC , route de Souesmes, 18330 Nancay, France \label{27}
 \and
Jodrell Bank Center for Astrophysics, School of Physics and Astronomy, The University of Manchester, Manchester M13 9PL, UK\label{28}
 \and
Astronomical Institute 'Anton Pannekoek', University of Amsterdam, Postbus 94249, 1090 GE Amsterdam, The Netherlands \label{29}
 \and
Max-Planck-Institut f\"{u}r Radioastronomie, Auf dem H\"ugel 69, 53121 Bonn, Germany \label{30}
 \and
Astronomisches Institut der Ruhr-Universit\"{a}t Bochum, Universitaetsstrasse 150, 44780 Bochum, Germany \label{31}
 \and
Astrophysics, University of Oxford, Denys Wilkinson Building, Keble Road, Oxford OX1 3RH \label{32}
 \and
Astro Space Center of the Lebedev Physical Institute, Profsoyuznaya str. 84/32, Moscow 117997, Russia \label{33}
 \and
Center for Information Technology (CIT), University of Groningen, The Netherlands \label{34}
 \and
Centre de Recherche Astrophysique de Lyon, Observatoire de Lyon, 9 av Charles Andr\'{e}, 69561 Saint Genis Laval Cedex, France \label{35}
\newpage
 \and
Fakult\"{a}t f\"{u}r Physik, Universit\"{a}t Bielefeld, Postfach 100131, D-33501, Bielefeld, Germany \label{36}
 \and
Department of Physics and Elelctronics, Rhodes University, PO Box 94, Grahamstown 6140, South Africa \label{37}
 \and
SKA South Africa, 3rd Floor, The Park, Park Road, Pinelands, 7405, South Africa \label{38}
 \and
Argelander-Institut f\"{u}r Astronomie, University of Bonn, Auf dem H\"{u}gel 71, 53121, Bonn, Germany \label{39}
}

\date{ Received /
		Accepted }

\abstract {The Sun is an active source of radio emission which is often associated with energetic phenomena such as solar flares and coronal mass ejections (CMEs). At low radio frequencies (<100 MHz), the Sun has not been imaged extensively because of the instrumental limitations of previous radio telescopes. } 
{Here, the combined high spatial, spectral and temporal resolution of the Low Frequency Array (LOFAR) was used to study solar Type III radio bursts at 30--90~MHz and their association with CMEs.}
{The Sun was imaged with 126 simultaneous tied-array beams within $\leq$5~\textit{R$_\sun$} of the solar centre. This method offers benefits over standard interferometric imaging since each beam produces high temporal ($\sim$83~ms) and spectral resolution (12.5~kHz) dynamic spectra at an array of spatial locations centred on the Sun. LOFAR's standard interferometric output is currently limited to one image per second. }
{Over a period of 30 minutes, multiple Type III radio bursts were observed, a number of which were found to be located at high altitudes ($\sim$4~\textit{R$_\sun$} from the solar center at 30~MHz) and to have non-radial trajectories. These bursts occurred at altitudes in excess of values predicted by 1D radial electron density models. The non-radial high altitude Type III bursts were found to be associated with the expanding flank of a CME. }
{The CME may have compressed neighbouring streamer plasma producing larger electron densities at high altitudes, while the non-radial burst trajectories can be explained by the deflection of radial magnetic fields as the CME expanded in the low corona.}


\maketitle

\section{Introduction}

{The Sun is an active star that produces large-scale events such as coronal mass ejections (CMEs) and solar flares. Radio emission is often associated with these events in the form of radio bursts. These bursts are classified into five main types \citep{wild63}. Type I bursts are short duration narrowband bursts associated with active regions \citep{mel75}. Type II bursts are slow frequency drifting radio emissions thought to be excited by shock waves travelling through the solar corona and they are associated with CMEs \citep{nel85}. Type III radio bursts are rapid frequency drifting bursts which can sometimes be followed by continuum emissions; these emissions are called Type V radio bursts \citep{wild63}. Type IV bursts are broad continuum emissions with rapidly varying time structures \citep{McL85}.}

{Type IIIs are rapidly varying bursts of radiation at metre wavelengths with durations of a few seconds. They were first described by \citet{wild50} who found that they had a characteristic frequency drift from high to low frequencies in solar dynamic spectra. More recently, Type IIIs have been observed at kHz frequencies \citep{kr10} up to frequencies of 8~GHz \citep{ma12}, although most occur at frequencies <150~MHz \citep{sa13}. In terms of drift rates, Type III bursts vary from approximately -1~MHz s$^{-1}$ at 20~MHz to -20~MHz s$^{-1}$ at 100~MHz \citep{ab90, mann02}, while their brightness temperatures can be up to $10^{12}$~K for coronal Type IIIs. High brightness temperature Type IIIs are generally associated with flaring activity on the Sun, however 90\% of the time Type III bursts occur in the absence of flares or CMEs \citep{du85}.}

\begin{figure*}[ht]
	\centering
	\includegraphics[ trim = 20px 0px 0px 0px, width =535px ]{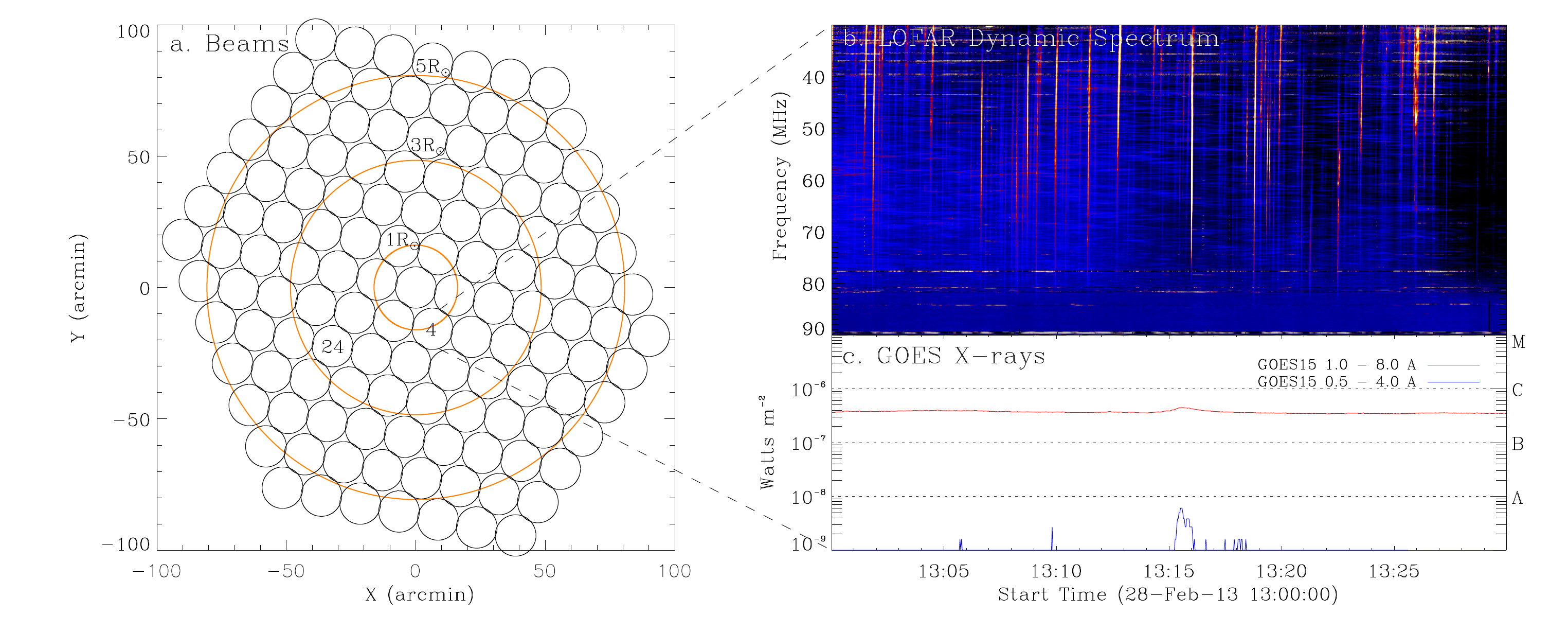}
	\caption{ \textbf{a)} Map of a LOFAR 126-beam tied-array pointing at the Sun. The center beam is pointing at the Sun center with successive beams around it in a honeycomb pattern to cover the station field-of-view (FOV) of $\sim$3.3\degr. The circles represent the beam size at a frequency of 45~MHz. The separation between beam centers is $\sim$$14\arcmin$.  \textbf{b)} 30-minute dynamic spectrum recorded by one of the LOFAR tied-array beams (Beam 4) indicated in panel a, on 2013 February 28. There are multiple Type III radio bursts in this observation. \textbf{c)} X-ray lightcurve measured by the GOES satellite during the time of the observation for two channels. The labels A, B, C and M represent solar flare classes based on the solar X-ray flux, with A being the smallest class of flares.}
	\label{fig:fig1}
\end{figure*}

{Type III bursts are considered to be the radio signature of electron beams travelling through the corona and into interplanetary space along open magnetic field lines \citep{lin74} and these electron beams are believed to originate from magnetic reconnection or shocks \citep{du00}. There are a number of theoretical explanations for Type III bursts, but it is commonly believed that, following acceleration, faster electrons outpace the slower ones to produce a bump-on-tail instability in their velocity distribution. This generates Langmuir (plasma) waves \citep{ro93} which are then converted into radio waves at the local plasma frequency ($f_p$) and its harmonic \citep[$2f_p$;][]{ba98}. Higher harmonics of the plasma frequency are associated with non-linear plasma processes and therefore this emission is a rare event \citep{zl98}. More recently, \citet{li13} have suggested that $f_p$ emission can be excited in a plasma where background particles have isotropic kappa velocity distributions (particle distributions with non-Maxwellian suprathermal tails decreasing as a power law of the velocity). Emission at the plasma frequency by kappa-distributed plasmas can reach observable intensity levels more easily than Maxwellian plasmas.}

{The spatial characteristics of Type III radio bursts have been studied in detail at specific frequency ranges: kHz up to $\sim$10~MHz and $>$150~MHz, leaving a gap at frequencies of $\sim$10--150~MHz. This was due to availability of radio instruments onboard the Ulysses \citep{sto92}, WIND \citep{bo95} and STEREO spacecraft \citep{ka05} and the Nan{\c c}ay Radioheliograph \citep[NRH;][]{ke97}. At low frequencies, \citet{re95} measured the trajectory of two Type III radio bursts at frequencies of kHz up to 1 MHz observed by the Ulysses spacecraft. They showed that they follow Parker spiral-like paths through interplanetary space up to $\sim$1 AU. More recent work by \citet{re09} and \citet{the10} with STEREO and WIND at frequencies $<$16 MHz used triangulation and ray-tracing techniques. \citet{re09} found that individual Type III bursts exhibit a wide beaming pattern approximately along the interplanetary magnetic field. At higher frequencies, some Type III radio bursts were imaged by the Culgoora Radioheliograph at 80~MHz \citep{wi67}. More recently, \citet{sa13} attempted a statistical study of almost 10000 Type III radio bursts observed by NRH over 7 years and found that at high frequencies (432~MHz), radio bursts are concentrated in two distinct bands around latitudes of 15\degr to 30\degr and -15\degr to -30\degr. The spatial characteristics of the low frequency range, 10--90 MHz, still remain unexplored with the location of the Type III sources being constrained by density models such as \citet{new61}, \citet{mann99} and \citet{zu14}. }

{Several low--frequency radio telescopes have been developed to observe solar radio activity, including interferometers, spectrometers and imaging-spectrometers. One of the earliest solar radio telescopes, the Culgoora Radioheliograph, started operations in 1968 at a frequency of 80~MHz, but discontinued in 1986 \citep{sh72}. The Nan{\c c}ay Decametric Array, operating since 1986, produces dynamic spectra at frequencies of 10--80~MHz \citep{bo80}. A recently upgraded radio telescope, UTR-2, produces dynamic spectra in the range 8–-32 MHz with much higher temporal and spectral resolution \citep{me11}. Recently, a range of low--frequency radio imaging arrays have been developed such as the Murchison Widefield Array \citep[MWA;][]{mwa13} and  the LOw Frequency ARray \citep[LOFAR;][]{lofar13}, that can offer new insight into the poorly explored low-frequency radio spectrum and into the propagation of Type III bursts. LOFAR represents a new milestone in low--frequency radio instrumentation operating at frequencies of 10--240~MHz. It features full polarisation and multi-beaming capabilities, with unprecedented sensitivity compared to previous radio telescopes due to the large number of LOFAR antennas.}

\begin{figure*}[ht]
	\centering
        \includegraphics[ trim = 90px 20px 90px 20px, width=350px, angle = 90 ]{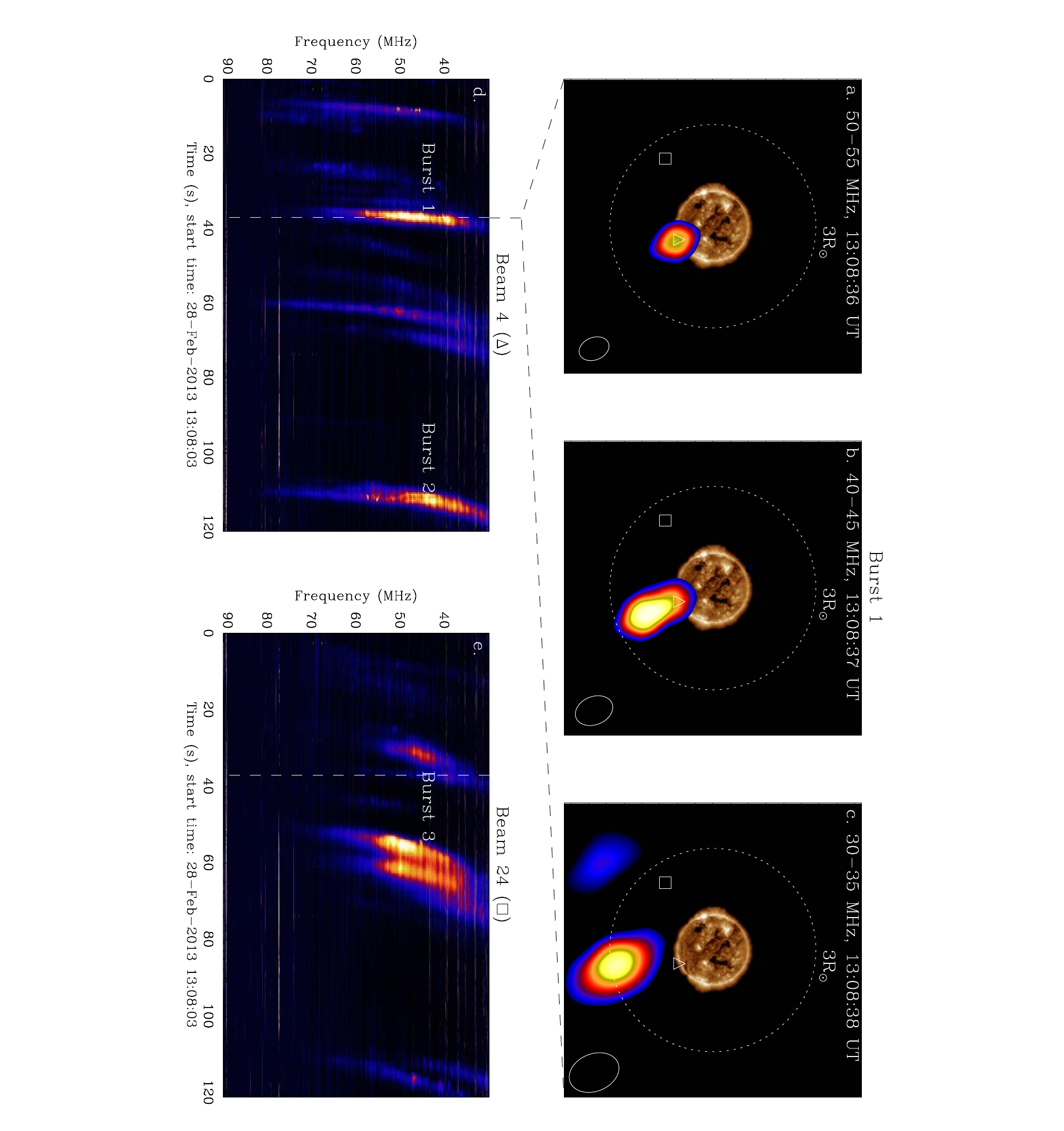}
	\label{fig:fig2}
	\caption{Tied-array images of a Type III radio burst at (a) 50--55~MHz (b)  40--45~MHz and (c) 30--35~MHz, separated by 1~s. The inset is an SDO/AIA (193~\AA) image of the EUV Sun at 13:08:18 UT. The dynamic spectra corresponding to two different beams, Beam 4 and Beam 24, are shown in panels d and e. The locations of Beams 4 and 24 are marked in panels (a), (b) and (c) with a triangle and square symbol, respectively. In panels (d) and (e), the white line identifies the timestamp in panel (b). The full evolution of the radio sources in the dynamic spectra (Bursts 1, 2 and 3) can be seen in the accompanying movie.}
\end{figure*}

{In this paper, we use LOFAR tied-array beams imaging and spectroscopy to study the spatial characteristics of over 30 Type III radio bursts several hours after the launch of two CMEs. These low--intensity bursts were detected due to LOFAR's increased sensitivity and can be studied with unprecedented temporal and spectral resolution. In Section 2.1 we give an overview of the LOFAR instrument, the observational mode used and the tied-array imaging method. In Section 2.2 we present methods of position and velocity estimations. In Section 3 we present our results which we compare to current density models of the solar corona, and then discuss these results in Section 4.}

\section{Observations and Data Analysis}

\subsection{Observations}

{On 2013 February 28, over 30 Type III radio bursts were observed by LOFAR. They were analysed in conjunction with extreme ultraviolet (EUV) images from the Solar Dynamics Observatory (SDO) and coronograph images from the Large Angle and Spectrometric COronagraph \citep[LASCO;][]{br95} onboard the Solar and Heliospheric Observatory (SOHO). These observations were made during times of low solar activity on the visible solar disk; however a CME was observed by LASCO originating behind the solar limb. EUV images were recorded by the Atmospheric Imaging Assembly \citep[AIA;][]{le12} onboard NASA's SDO mission which provides an unprecedented view of the solar corona at multiple wavelengths, imaging the Sun up to $\sim$1.1~\textit{R$_\sun$} from the solar centre. In this paper, the LASCO/C3 coronagraph was used to image the Sun at radial distances of $\sim$3--32~\textit{R$_\sun$}.}

{LOFAR is a new-generation radio interferometric array constructed by the Netherlands Institute for Radio Astronomy (ASTRON). LOFAR consists of thousands of dipole antennas distributed in 24 core stations and 16 remote stations throughout the Netherlands and 8 international stations across Europe. There are two distinct antenna types: the Low Band Antennas (LBAs) which operate at frequencies of 10--90~MHz and the High Band Antennas (HBAs) which operate at 110--240~MHz \citep{lofar13}. The large number of LOFAR antennas results in a large collecting area of 35000~m$^2$ at a frequency of 30~MHz. LOFAR is capable of improving the sensitivity and angular resolutions of solar observations by 2 orders of magnitude compared to previous radio telescopes \citep{mann11}.}

\begin{table*}[t]
\centering
\caption{\label{table1} LOFAR heights and velocities derived from tied-array imaging and predictions for these values for various density models of the solar corona for the Type III radio bursts labelled 1, 2 and 3 in Figure 2. We assume harmonic emission for all the Type III radio bursts ($n=2$ in Equation 1). }
	\begin{tabular}{ccccc} 
	\hline\hline
	                                 & LOFAR Tied-array Imaging   & Mann Model  & Newkirk Model & Zucca Model \\ \hline \hline
	& & Burst 1  \\ \hline 	
	Start Distance ($R_\sun$) & $1.18\pm0.7$       & $1.37$  & $1.57$    & $1.50$\\ 
	End Distance ($R_\sun$) & $2.90\pm1.2$       & $1.72$  & $2.16$    & $1.95$\\ 
	Velocity ($c$)             & $0.51\pm0.18$    & {0.11}             & {0.20}               & {0.14}\\ \hline
	& & Burst 2  \\ \hline 
	Start Distance ($R_\sun$) & $2.64\pm0.7$       & $1.30$  & $1.60$    & $1.55$\\ 
	End Distance ($R_\sun$) & $4.06\pm1.2$       & $1.74$  & $2.41$    & $1.96$\\ 
	Velocity ($c$)             & $0.27\pm.0.11$    & {0.05}             & {0.09}               & {0.05}\\ \hline
	& & Burst 3  \\ \hline 
	Start Distance ($R_\sun$) & $1.16\pm0.7$       & $1.39$  & $1.81$    & $1.60$\\ 
	End Distance ($R_\sun$) & $2.77\pm1.2$       & $1.70$  & $2.40$    & $1.98$\\ 
	Velocity ($c$)             & $0.58\pm0.17$     & {0.07}             & {0.12}               & {0.08}\\ \hline

	\end{tabular}
\end{table*}

{In this paper we have taken observations using the LBAs of LOFAR core stations located in Exloo, Netherlands where the maximum baseline is $\sim$2~km. All 24 LOFAR core stations have been used for these observations. Instead of producing interferometric visibilities, the data analysed here were acquired using one of LOFAR's beam-formed modes \citep{sta11, lofar13}. This was used to combine the data from the LOFAR core stations into multiple ``tied-array beams''. The advantage of using tied-array beams over interferometric imaging is that the data acquired from each beam were used to produce dynamic spectra with a high temporal and spectral resolution (12.5~kHz; $\sim$83~ms), which is not possible to obtain using standard LOFAR interferometric observations. In addition to the high--resolution dynamic spectra, the tied-array beams offer the possibility of imaging solar radio bursts with very high temporal resolution (that of the dynamic spectra). For comparison, LOFAR's standard interferometric output is limited to approximately one image per second.  }

{126 simultaneous beams were used to observe the Sun on 2013 February 28 for a period of 30 minutes at 13:00--13:30 UT (Figure 1a). The range of the full-width half maxima (FWHM) of the beams ranged between 7\arcmin~at 90~MHz to 21\arcmin~at 30~MHz. The separation between beam centres is $\sim$$14\arcmin$ and as a result the beams were touching at a frequency of 45~MHz as seen in Figure 1a and began to overlap at lower frequencies. Each beam recorded a high frequency and high time resolution (12.5~kHz; $\sim$83~ms) dynamic spectrum (Figure 1b). Due to radio frequency interference (RFI) below 30~MHz, we chose the frequency range of the LOFAR dynamic spectra to be 30--90 MHz as opposed to the full frequency range of the LBAs. No calibrator was observed, therefore intensity values throughout this analysis are expressed in arbitrary units. }

\subsection{Data Analysis}

{The intensity value in each of the tied-array beams at a given frequency and time was extracted from each of the 126 dynamic spectra and plotted as `macro-pixels' onto the tied-array map in Figure 1a. We produced images of radio bursts by interpolating these intensity points to the nearest beam. In order to reduce spectral noise in the dynamic spectra, the data were averaged over frequency. The optimal averaging frequency bandwidth was found to be 5~MHz, resulting in images corresponding to a 5~MHz frequency bin (eg. 30--35~MHz,  35--40~MHz, 40--45~MHz). A time sampling of 1~s was chosen in order to observe the full propagation of the Type III bursts and the data were therefore averaged in time over 1~s.}

\begin{figure*}[ht]
	\centering
	\includegraphics[ trim = 170px 160px 160px 90px, clip, width=250px, angle = -90]{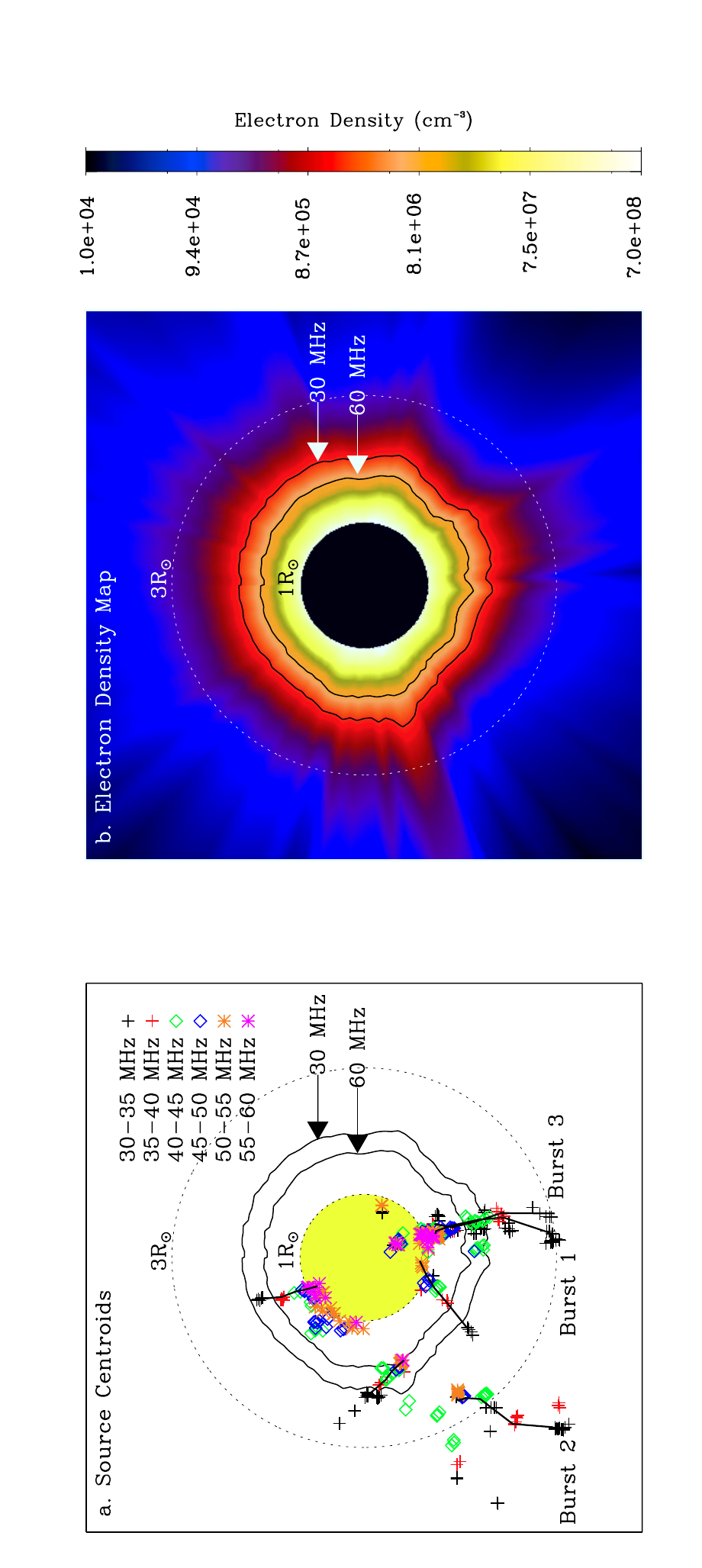}
	\label{fig:fig3}
	\caption{ \textbf{a)} Centroids of Type III radio burst emission sources observed during a 30 minute period at frequencies between 30 and 60~MHz sampled in 5~MHz--wide frequency bins. The solid lines represent the motion of some of the Type III radio bursts within each individual group. \textbf{b)} Electron density map of the solar corona \citep{zu14}. The solid contours are plotted for the 30~MHz and 60~MHz emission sites assuming harmonic plasma emission in this density map.}
\end{figure*}

{This processing method resulted in a sequence of images of scattered points which have an intensity value corresponding to the position of each individual beam. As can be seen in Figure 1a, the raw data has a honeycomb pattern which is the pattern of tied-array beams. In order to easily process the data, the scattered intensity points were interpolated with a smooth quintic surface to produce a regularly gridded array. After interpolating and regridding the data, a regular 2D array is obtained which can be used to image the radio sources as seen in Figures 2a, 2b and 2c as opposed to a honeycomb-shaped image.}

{The spatial information of each individual radio source was extracted from these maps for a few frequency ranges (30--35~MHz, 40--45~MHz etc.). The centroids of each source were transformed into distances in solar radii starting from the solar center. The centroids were then used to produce distance-frequency plots for comparison with density models of the solar corona (Figure 3) and also distance-time plots for velocity estimations of the Type III radio sources (Figures 4 and 6).}

{The elevation of the Sun at the time of the observation was $\sim$30\degr. Due to the low elevation of the Sun, we have investigated the effects of the ionosphere in calculating the positions of Type IIIs in our observation. This was done by looking at the quiet Sun position in the tied-array images. The maximum intensity point of the quiet Sun was found to be consistent with the center of the tied-array beams throughout the observations. The maximum variation observed was $\sim$0.5~$R_\sun$.  This was tested for frequencies between 30--90~MHz. Some variation was expected due to the imaging method used that relies on interpolation of intensity points to the nearest beams with a spatial separation between the beams of $\sim$1~$R_\sun$. We have also calculated spatial corrections due to the ionosphere taking into account the position of the Sun at the time of the observation \citep{st82, me96}. The maximum corrections found were $\sim$2\arcmin~which are significantly smaller compared to the FWHM of the beams which can go up to 21\arcmin~(1.3~\textit{R$_\sun$}) at 30 MHz. As a result no ionospheric corrections were applied to this analysis. }

\section{Results}

\subsection{Type III Spectral Characteristics}

{Figure 1b shows the dynamic spectrum recorded for 30 minutes on 2013 February 28 at 13:00--13:30 UT by one of the tied-array beams (Beam 4), while Figure 1c shows the GOES (Geostationary Operational Environmental Satellite) X-ray flux during this period. No significant X-ray activity was observed on 2013 February 28 before or during the LOFAR observation; there were minor B class flares observed after the time of the LOFAR observation. Although there were no major X-ray flares at the time, a number of CMEs were observed by coronagraphs on STEREO and SOHO in the hours preceding the LOFAR observations. The CMEs appear to have originated from the far side of the Sun relative to Earth. A Type III group or storm was observed during the time interval the observations were taken in. } 

{A total of 32 Type III radio bursts have been observed during a period of only 30 minutes, after an intensity threshold was set to distinguish the strongest radio bursts and avoid noise (Figure 1b). A zoom in on three particularly strong Type III bursts is also shown in Figures~2d and 2e. The frequency drift rates of the 32 bursts analysed here vary between -2 to -17~MHz~s$^{-1}$  with an average of -7~MHz~s$^{-1}$ at 30--60~MHz, which are comparable with the findings of \citet{mann02}, who reported a mean value of -11~MHz~s$^{-1}$ at a frequency range of 40--70~MHz.}

{As the radio bursts  travel outwards through the corona, the frequency of emission of these radio bursts  decreases with distance. Assuming plasma emission, the emission frequency of Type III radio bursts can be calculated using
\begin{align}
\label{eq:1}
f = n f_p = n~C\sqrt{N_e} ~\textrm{Hz} .
\end{align} 
The emission frequency, $f$, is given by the local plasma frequency, $f_p$, multiplied by the harmonic number $n$ and is directly proportional to the square root of the electron density, $N_e$, in cm$^{-3}$, where $C = 8980$~Hz~cm$^{3/2}$ is the constant of proportionality. The electron density decreases with altitude in the corona and, therefore, the frequency also decreases with altitude. It is possible to estimate the radial velocity of the Type IIIs using the frequency drift rate, $df/dt$, of the Type III bursts calculated according to the equation
\begin{align}
\label{eq:1}
v = \frac{2\sqrt{N_e}}{C} \left({\frac{dN_e}{dr}}\right)^{-1} \frac{df}{dt} .
\end{align} }

\begin{figure}[ht]
	\includegraphics[ trim = 0px 0px 20px 0px, width = 188px, angle = 90 ]{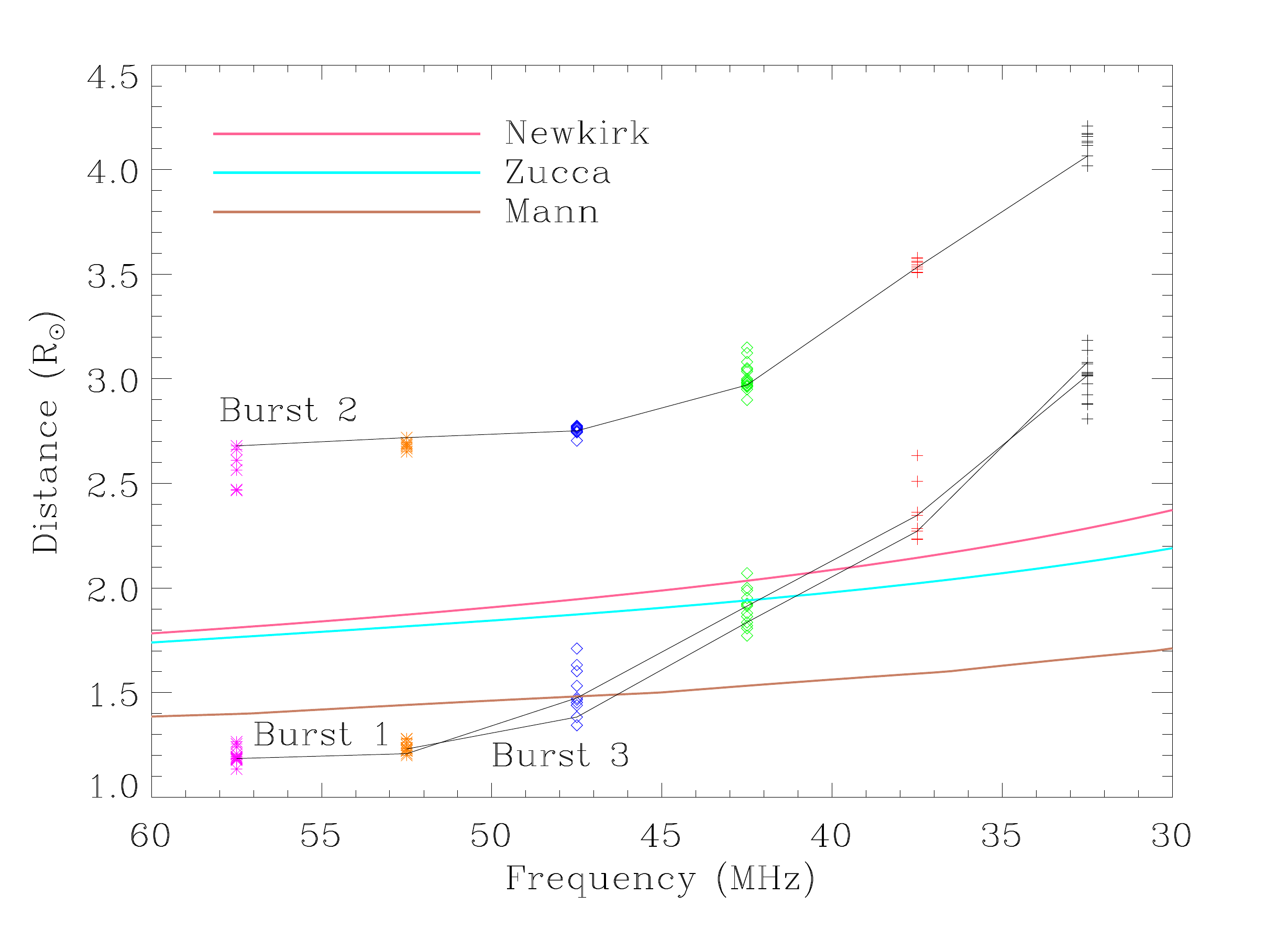}
	\caption{Height-frequency plot of the Type III radio bursts observed in Figure 2d and 2e. The trajectories of Bursts 1, 2 and 3 from Figures 2d and 2e are shown by the black solid lines. The coloured solid lines represent electron density models of the solar corona assuming harmonic plasma emission.}
	\label{fig:fig4}
\end{figure}

{In Equation 2, the unknown is generally the local plasma density, $N_e$. Since the electron density is assumed to decrease radially in the corona it is necessary to consider a density model, $N_e=N_e(r)$, to obtain the density at a specific height in the corona. Here, we have used the radial electron density models of \citet{new61} and \citet{mann99} and the time-dependent density model of \citet{zu14} to estimate the altitude and velocity of these bursts which are given in Table~1 and Figures 4 and 6 assuming harmonic plasma emission ($n=2$ in Equation 1). We have made this assumption because the free-free absorption for harmonic radiation is less severe and it can escape emission sites easier than fundamental radiation \citep{ba98}. In addition, the sources in this observation are found to be at higher altitudes where harmonic emission is expected to occur based on the local plasma density.  }

\begin{figure}[ht]
	\centering
	\includegraphics[ trim = 70px 100px 60px 150px,  width=250px, angle = 90 ]{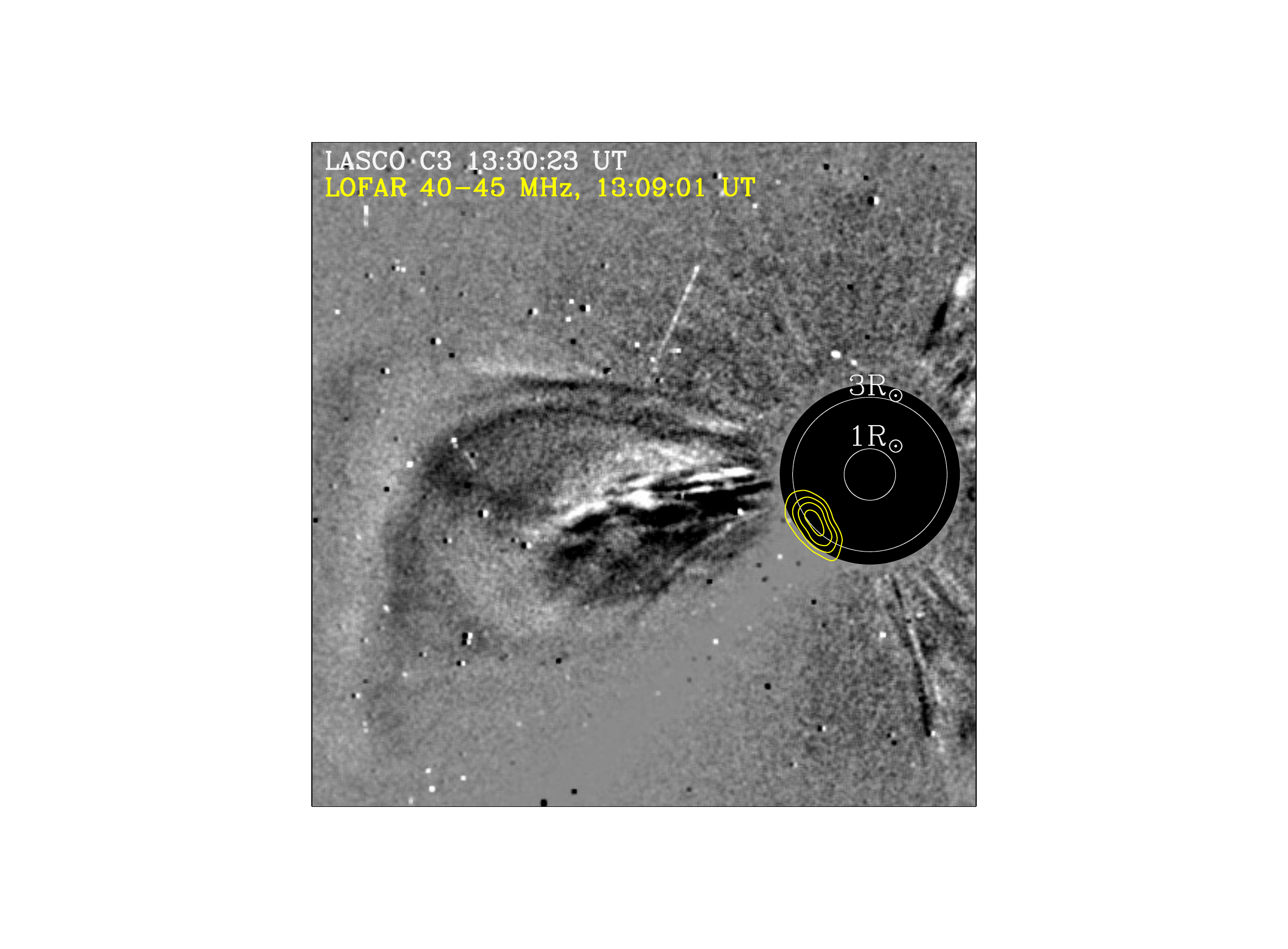}
	\label{fig:fig5}
	\caption{LASCO C3 running difference image of a CME on 28 February 2013 at 13:30:23 UT. The high altitude Type III burst at 13:09:01 UT (Burst 2) is shown in yellow contours. }
\end{figure}

\subsection{Type III Spatial Characteristics}

{Tied-array images of the Type III radio sources were produced at frequencies between 30 and 60 MHz, in 5 MHz--wide frequency bins. The images of a radio source at 50--55 MHz, 40--45 MHz and 30--35~MHz, respectively are shown in Figures 2a, 2b and 2c together with an AIA 193~\AA\ passband image from 13:08:18~UT. These radio images are 1~s apart in order to show the motion of the Type III burst. Figures 2d and 2e show the dynamic spectra of two different beams, Beam 4 and Beam 24, denoted by the triangle and square symbol, respectively, in the panels above. These two dynamic spectra were chosen to show the variety of Type III radio bursts occurring at different positions relative to the Sun.  The Type III labelled Burst 1 in the dynamic spectra is imaged in Figures 2a, 2b and 2c and the dashed white line in the dynamic spectra matches the timestamp in Figure 2b. The accompanying movie shows the full evolution of Bursts 1, 2 and 3 over a period of 2 minutes. }

{The propagation of the 32 Type III bursts observed is shown in Figure 3a. Here, the centroids of all Type III radio bursts observed during the 30 minute time period are shown over a frequency range of 30 to 60 MHz in 5 MHz--wide frequency bins. The motion of Bursts 1, 2 and 3 from Figures 2d and 2e are shown by the black solid lines. An electron density map of the solar corona combining both observations and models is shown in Figure 3b. Electron densities in this map were estimated from the differential emission measure (DEM) derived using SDO/AIA's six coronal filters for the height range 1--1.3~\textit{R$_\sun$} and using polarised brightness SOHO/LASCO for 2.5--5~\textit{R$_\sun$}. For the height range 1.3--2.5~\textit{R$_\sun$} a combined plane-parallel and spherically-symmetric model was employed \citep[See][for further details]{zu14}. The density map was produced for the time period of the LOFAR observations. However, the time evolution of the map is limited by the cadence of the LASCO C2 data ($\sim$20~minutes). While the map gives an accurate description of the electron density situation around the time of the observations, it cannot be made contemporaneous with the individual tracks of the radio bursts. }

{Figures 3a and 3b also show the location of the contours of the electron densities necessary for 30~MHz and 60~MHz harmonic plasma emission ($n=2$ in Equation 1). Using Equation 1, the electron densities required for radio emission at these frequencies are 2.8$\times$$10^6$ cm$^{-3}$ and 1.1$\times$$10^7$ cm$^{-3}$, respectively. The positions of the Type III centroids lie on either side of these contours in Figure 3a. This can be a consequence of the spatial resolution of the imaging technique used which is lower than that of the density map or it may be due to narrow-high density streamers not visible in the density map due to the limited spatial resolution of the LASCO data. However, Burst 2 was found at even higher altitudes which correspond to much lower electron densities in Figure 3b. }

{The motion of Bursts 1, 2 and 3 are shown in Figure 4 in a distance-frequency plot. The solid lines in Figure 4 represent the three electron density models mentioned above for comparison with the motion of the bursts. The density model curves are plotted assuming harmonic plasma emission ($n=2$ in Equation 1) as well as a radial electron density profile through the corona. There is no emission between 55--60~MHz for Burst 3 because the intensity falls below the intensity threshold used. The slope of the motion of Bursts 1 and 3 observed by LOFAR is steeper than the slope of the density model curves. This implies that the electron density gradient in the corona is steeper than that	 assumed by the density models. Burst 2 occurs at altitudes significantly higher than Bursts 1 and 3 and the predictions of the density models. Burst 2 therefore originates at a higher altitude in the corona and the slope of the motion curve for Burst 2 also has a steeper profile. }

\subsection{High Altitude Type III Bursts}

{Burst 2 has a non-radial trajectory as seen in Figure 3a and its start altitude is significantly higher than the predictions of the three density models in Figure 4. There is also no density enhancement in Figure 3b in the direction of Burst 2. After investigating coronograph images, in particular LASCO C3 (Figure 5), a CME was observed propagating away from the Sun and expanding during the time the radio bursts occur (image at 13:30:23 UT). The CME is plotted in Figure 5 together with the LOFAR 40--45~MHz contour (yellow) of the high altitude Type III burst (Burst 2) at 13:09:01 UT. There are no features in the LASCO image in the direction radial to the Type III radio burst due to the pylon of the occulting disk of LASCO C3. However, the position of the southern leg of the CME in Figure 5 is cospatial with the position of the Type III radio burst contours. The CME was the only large-scale change in the solar corona occuring at the time of the observation that would account for an increase in density up to $\sim$4~\textit{R$_\sun$}. As can be seen in the electron density map in Figure 4b, the streamers present did not have sufficiently high densities to account for this emission. }

\begin{figure}[ht]
	\includegraphics[ width = 250px ]{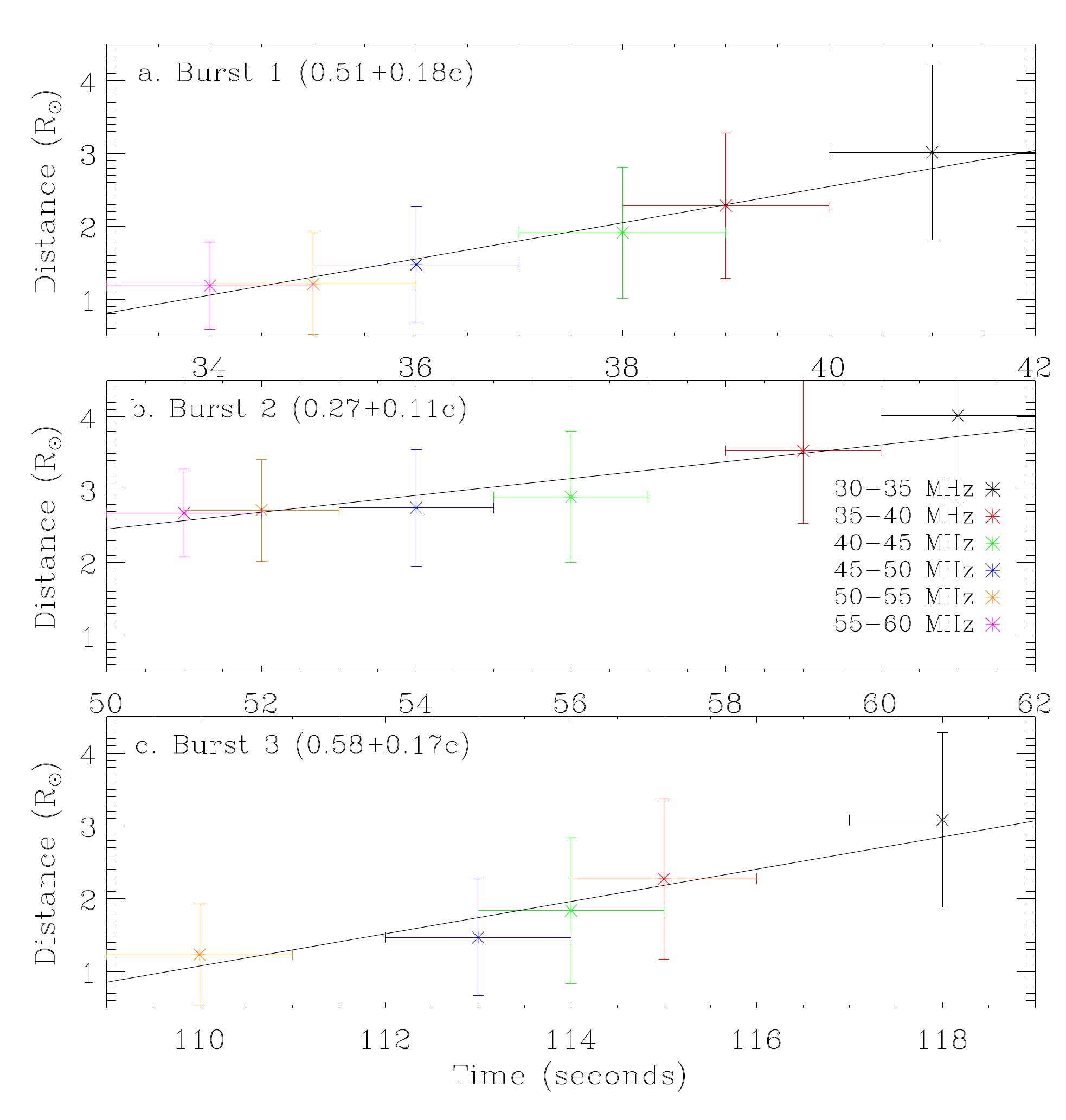}
	\caption{Radial motion of Bursts 1, 2 and 3. The velocities given on the left of each panel were calculated from the slope of the fit. The colours correspond to different frequency bins from 30 to 60~MHz. The error bars in the three plots are calculated based on the FWHM of the beams at each frequency.}
	\label{fig:fig6}
\end{figure}

{At a frequency of 32.5~MHz, a Type III radio burst would require an electron density of 3.3$\times10^6$ cm$^{-3}$ (see Equation 1) while in our density map the local plasma densities are of the order of $10^5$~cm$^{-3}$. There are two ways in which the local plasma can have high electron densities at 4~\textit{$R_\sun$}: the direct passage of the CME compresses the local plasma around it or the CME expanding against a coronal streamer which compresses it increasing the electron density inside the streamer. The CME may also deflect the streamer and as a result deflect the existing radial open magnetic field lines to non-radial directions. Since streamer activity was observed before and after the CME occurred, a dense streamer was possibly further compressed by the CME so that electron densities could have been high enough to explain these high altitude radio bursts.}

\subsection{Type III Kinematics}

{The centroids of Bursts 1, 2 and 3, which can be seen in the accompanying movie, were extracted in order to produce height-time plots as seen in Figures 6a, 6b and 6c. The colours in Figure 6 correspond to radio emission between 30 and 60~MHz, averaged in 5~MHz--wide frequency bins. There is no emission between 55--60~MHz for Burst 3 because the intensity falls below the intensity threshold used. The slope of each plot was used to estimate the velocity of propagation of the bursts. The error bars in the three plots represent the largest calculated errors which occur due to the large size of our tied-array beams at these low frequencies (this is given by the FWHM of the beams at each frequency). }

{Estimated distances and velocities of Bursts 1, 2 and 3 from LOFAR imaging and various density models are also shown in Table 1 for direct comparison with these models. The velocities of Bursts 1, 2 and 3 (0.58$\pm$0.17~\textit{c}, 0.27$\pm$0.11~\textit{c} and 0.51$\pm$0.18~\textit{c}, respectively) are generally higher than the standard velocities of Type III radio bursts of $\sim$0.05--0.30~c \citep{du87, lin81, lin86}. The motion of the Type III sources contradicts the predictions of various density models using frequency drift rates as seen in Table 1. \citet{wild59} found higher velocities from interferometry in a number of Type III radio bursts studied, exceeding $\sim$0.3~\textit{c}, with an average velocity of 0.45~\textit{c}. However, the high velocities could also be a consequence of the low spatial resolution at low radio frequencies of both observations but this is unlikely for most of the bursts as their movement is larger than the extent of the beam. In this observation, the density models used are not a good match for the motion of the Type III radio bursts observed since they predict lower altitudes and velocities compared to LOFAR tied-array imaging.}

\section{Discussion and Conclusion}

{Using a novel observing mode, the LOFAR tied-array beams, we extracted the spatial information of multiple Type III radio bursts and related it to high-time and frequency resolution dynamic spectra. Our observations show a number of Type III radio bursts being emitted at high altitudes (30--35~MHz bursts at 4~\textit{R$_\sun$} from the solar center) which is not believed to occur due to the low electron density of the surrounding plasma in a quiet solar corona. Our observations also show that not all Type III radio bursts follow a radial trajectory from the center of the Sun as previously assumed by electron density models of the corona. It is therefore necessary to consider a 2D electron density map \citep{zu14} for Type III studies. However, such density maps are based on polarised brightness from LASCO data in the range 2.5--5~\textit{R$_\sun$} and narrow streamers cannot be taken into account due to limited resolution. Since ionospheric effects did not play an important role into shifting the source positions, we investigated the relation of the high altitude Type III bursts with large-scale changes in the corona that have an impact on the local plasma density such as coronal streamers and CMEs. A CME was indeed observed by LASCO and the southern leg correlates spatially with the location of the Type III bursts. Since streamer activity was observed before and after the CME occurred, we believe a dense streamer was compressed by the CME. This resulted in a higher density region inside the streamer necessary for high altitude plasma emission at 30~MHz assuming electrons were accelerated to produce the Type III radio bursts. }

{The velocities calculated from the motion of the sources are generally higher than Type III velocities reported in the literature of $\sim$0.05--0.30~\textit{c} \citep{du87, lin81, lin86}. The velocities were also calculated using various density models applied to the frequency drift rate in the dynamic spectrum (Table 1). The velocities varied with density model but they were lower by up to a factor of $\sim$4 than the velocities calculated from LOFAR tied-array imaging. This comes as no surprise since it was already shown that the density models used do not represent the situation well especially in the case of the CME associated Type III bursts. The high velocities of tied-array imaging could be argued to be a consequence of the low spatial resolution of the beams. However the displacement of the Type IIIs exceeds the beam size and errors due to the FWHM of the beams were included in the velocity calculations. Therefore the electron density models used are not a good match for our observations.  }

{The high altitude Type III bursts were shown to be associated with the passage of a CME. \citet{ca13} found particle acceleration associated with a CME shock in the solar corona which produced a herringbone pattern on the backbone of a Type II radio burst. Herringbones are indicative of electron beams travelling away from the shock and they are similar in appearance and velocity to Type III bursts \citep[][found herringbone generating electrons to have a velocity of $\sim$0.15~\textit{c}]{ca13}. We applied the same analysis to our Type III bursts; however, the CME observed on 2013 February 28 was a slow CME with a radial velocity of $\sim$250~km~s$^{-1}$. The Alfv\`{e}n speed at 4~\textit{R$_\sun$} was estimated to be $\sim$220~km~s$^{-1}$. Since the CME speed was not very high compared to the Alfv\`{e}n speed, the shock produced was weak and therefore may not have been very effective at accelerating particles to velocities $>$0.1~\textit{c} which is the velocity of the Type III producing electrons.}

{An alternative scenario that may explain particle acceleration is magnetic reconnection along the flank of the CME. For example, \citet{be10} found multiple reconnection sites associated with a CME expanding against the surrounding coronal streamers which they verified with numerical simulations. In our observations, assuming the southern leg of the CME is in the vicinity of an open magnetic field line, electrons resulting from these side reconnections can escape and produce the high altitude Type III radio bursts we see, inside the denser plasma of a compressed streamer. The non-radial trajectory of the Type III bursts can also be explained by the deflection of a coronal streamer by the CME.}

{The tied-array beams prove promising in detecting radio bursts with very high frequency and temporal resolution. In addition to this, tied-array beam observations can measure the positions of the emission sources of the radio bursts. The accuracy in these positions is only limited by the beam size since ionospheric effects were not significant during this observation. The beams size can be improved by increasing the baseline in future observations. In principle, remote stations could be added to the full LOFAR core, however this is not yet possible. An adequate beam size for the study of the sources of Type III radio bursts would be approximately half of the current beam size (3.5\arcmin~at 90~MHz to 10\arcmin~at 30~MHz). Ionospheric effects may become more significant for smaller spatial scales. In the future, the possibility of making simultaneous observations using both LBAs and HBAs will be beneficial for full spectral analysis of these radio bursts. This analysis can also be applied to different types of bursts within the LOFAR spectral range.}

\begin{acknowledgements}

{This work has been supported by a Government of Ireland studentship from the Irish Research Council (IRC), the Non-Foundation Scholarship awarded by Trinity College Dublin and the Innovation Academy. We would also like to acknowledge the IRC New Frontiers Funding for supporting LOFAR related travel. Chiara Ferrari acknowledges financial support by the {\it “Agence
Nationale de la Recherche”} through grant ANR-09-JCJC-0001-01. We would finally like to acknowledge the LOFAR telescope which is made possible by the hard work of dozens of engineers, technicians, developers, observers and support scientists. LOFAR, the Low Frequency Array designed and constructed by ASTRON, has facilities in several countries, that are owned by various parties (each with their own funding sources), and that are collectively operated by the International LOFAR Telescope (ILT) foundation under a joint scientific policy.}

\end{acknowledgements}

\bibliographystyle{aa} 
\bibliography{morosan_AA_20140331} 

\end{document}